\documentclass[aps,preprintnumbers, superscriptaddress, amsmath, showpacs]{revtex4}
\usepackage{epsfig}
\usepackage{psfrag}

\usepackage{graphicx}
\usepackage{dcolumn}
\usepackage{bm}

\makeatletter
\renewcommand \thesection {\@arabic\c@section}
\makeatother
\makeatletter 
\@addtoreset{equation}{section}
\makeatother  

\begin{document}


\title{Mass Effect in Single Flavor Color
Superconductivity}

\author{Ping-ping Wu}
 \email{wupingping@iopp. ccnu. edu. cn}
\affiliation{Institute of Particle Physics,  Huazhong Normal
University, Wuhan,  430079,  China} \affiliation{The Key
Laboratory of Quark and Lepton Physics(HZNU), Ministry of
Education, Wuhan 430079, China}
\author{ De-fu  Hou}%
 \email{hdf@iopp. ccnu. edu. cn}
 \affiliation{Institute of Particle Physics,  Huazhong Normal University,
Wuhan,  430079,  China}

\author{Hai-cang Ren}
\email{ren@mail.rockefeller.edu}  \affiliation{Institute of
Particle Physics, Huazhong Normal University, Wuhan, 430079,
China} \affiliation{Physics Department, The Rockefeller
University, 1230 York Avenue,  New York,  NY 10021-6399}

\begin{abstract}

The nonzero strange quark mass effect in different types of single
flavor color superconductivity and the phase diagram in a magnetic
field are studied. We have obtained simple analytical forms of the
quasi-particle energies for an arbitrary mass and explored the
mass correction to the pressure and the transition temperature. It
is found that the mass reduces the pressure and transition
temperature of strange quarks, but it doesn't change the ranking
$P_n<P_{\rm A}<P_{\rm polar}<P_{\rm planar}<P_{\rm CSL}$ of the
pressure for the four canonical single flavor phases. The phase
diagram with magnetic field and temperature for a system of three
flavors is obtained for two different values of the strange quark
mass. The changes from the one obtained previously under the
approximation of massless strange quarks are examined.
\end{abstract}

\maketitle

\section{Introduction}\label{sec:level1}

Quark matter at sufficiently high baryon density and low
temperature becomes a color superconductor(CSC) \cite{MAKT}. CSC
is characterized by a diquark condensate, which is analogous to
the Cooper pair in an ordinary superconductor, but the structure
of the condensate is much richer because quarks have the
nonabelian color and flavor charges.

The structure of the CSC states depends sensitively on the number
of quark flavors and their masses
\cite{structure,description,symmetry,critical}. For very high
baryon density, where the masses of u, d and s quarks can be
ignored, the ground state is in the color-flavor-locked(CFL) phase
\cite{cfl}, where quarks of different flavors pair. The situation becomes more complicated in moderate
density because of the strange quark mass, $\beta$ equilibrium and
the charge neutrality conditions. A substantial Fermi momentum
mismatch among different quark flavors is introduced and thereby
reduces the available phase space for the cross-flavor pairing, such
as CFL. Different exotic scenarios for cross-flavor pairing
proposed in the literature (gapless CSC, LOFF CSC etc.) either run
into various instabilities \cite{mcfl,mcf2,chromomagnetic} or reduce
significantly the condensation energy. This makes the single
flavor pairing, which is free from the Fermi momentum mismatch and the instabilities, a
competing alternative even though the pairing force here is
expected to be weaker. There are a number of different paring
states. The ones frequently discussed in the literature include
the spherical color-spin-lock(CSL) and nonspherical planar, polar
and A \cite{SWR,T,equality}. Here, the adjective
"spherical/nonspherical" refers to the symmetry of the order
parameter under a space rotation. The CSL pairing is energetically most favored
in the absence of a magnetic field, but the situation changed when
a magnetic field is applied. The single flavor color
superconductivity may be realized in the interior of a compact
star during the later stage of its life, where a magnetic field is
present.

The presence of a magnetic field in the interior of a compact star
\cite{compact} will offset the energy balance among the four
canonical single flavor pairings. The spherical CSL phase has an
electromagnetic Meissner effect \cite{SWR}, but nonspherical
phases: polar, A and planar phases do not. So if a quark matter of
single flavor parings cools down through the critical temperature
in a magnetic filed, forming CSL state will cost extra work to
exclude magnetic fluxes from the bulk. Therefore, the magnetic
contribution to the free energy may favor the nonspherical states.
In a previous work, we have explored the consequences of the
absence of the electromagnetic Meissner effect in a nonspherical
CSC phase of single flavor pairing \cite{magne} and have obtained
the phase diagram with respect to the magnetic field and the
temperature. We found that under the plausible magnitude of the
the magnetic field inside a compact star, the most favored state
is not always CSL and nonspherical pairing states may show up. For
the sake of simplicity, we considered both the infinitely massive
limit and the massless limit strange quarks in \cite{magne}. The
former limit is unrealistic given the typical chemical potential
$\mu$ around 500MeV, the latter requires the mass of strange
quarks, $m_s$, to be much lower than quark chemical potential. On
the other hand, $m_s$ has to be sufficiently large in order to win
the competition with exotic cross-flavor pairings such as gapless
CSC and LOFF. Both requirements may be compromised marginally for
the value of $m_s$ in vacuum ($\sim$ 150MeV) but will be
problematic when the value of $m_s$ in medium becomes comparable
with $\mu$ as was suggested by some numerical works such as
\cite{mass}. All these concerns warrant a systematic treatment of
the single flavor pairing with an arbitrary quark mass. So we did
in this paper.

In the present work, we shall give a detailed investigation of the
phase structure. For this purpose, we formulate the single flavor
CSC for an nonzero quark mass in terms of the same
NJL(Nambu-Jona-Lasinio)-like effective action employed in
\cite{magne} and introduce the mean-field approximation for an
arbitrary mass in section 2. Unlike the ultra-relativistic limit,
where the cross-helicity(transverse) pairing dominates, the
nonzero quark mass couples the cross helicity pairing channel and
the equal-helicity(longitudinal) pairing channel and thereby
complicates the gap matrix underlying the excitation spectrum.
Fortunately, as will be shown in section 3, the gap matrix for an
arbitrary mass can still be diagonalized analytically for all four
canonical phases and our results interpolate both the
ultra-relativistic limit and the non-relativistic limit in the
literature. The ranking of the condensation energy in the massless
limit remains intact when a nonzero quark mass is switched on. In
section 4 we generalize our analysis in \cite{magne} of a
three-flavor quark matter beyond the ultra-relativistic limit.
Because the transition temperature of the nonzero $m_s$ strange
quark paring is reduced, phase diagram with respect to temperature
and magnetic field contains a region where only u and d flavors
condensate. The size of this region is tiny for $m_s\sim 150$MeV
but cannot be ignored for $m_s\sim\mu$. Finally, we summarize our
results and remark on some open issues in section 5. Throughout
the paper, we shall assume zero masses for u and d quarks as we
did in \cite{magne}. All gamma matrices are hermitian according to
our notation.

\section{The Hamiltonian under mean field approximation}

In this section and the next one, we shall formulate the single flavor
Copper pairing with an arbitrary quark mass.
The Lagrangian density
of the NJL-like effective action reads \cite{blaschke}:
\begin{eqnarray}
{\cal L}=\bar\psi(-\gamma_\nu\partial_\nu+m+\mu\gamma_4)\psi-G\bar\psi\gamma_\nu
T^l\psi\bar\psi\gamma_\nu T^l\psi \label{lag}
\end{eqnarray}
where $T^l=\frac{1}{2}\lambda^l$ with $\lambda^l$ the $l$th
Gell-Mann matrix, $m$ is the quark mass and $\mu$ is the chemical
potential. We set the effective coupling $G>0$, in accordance with
the interaction mediated by one-gluon exchange at high density and
that mediated by instantons for intermediate density. The
corresponding Hamiltonian is
\begin{equation}
H=\int
d^3\mathbf{r}\Big[\bar\psi(\mathbf{\gamma}\cdot\mathbf{\nabla}+m-\mu\gamma_4)\psi+G\bar\psi{\gamma_{\nu}}
T^l\psi\bar\psi{\gamma_\nu}{T^l}\psi\Big]\label{lag1}
\end{equation}
Like QCD Lagrangian, the diquark scattering in (\ref{lag})
conserves the eigenvalues of $\gamma_5$ of each quark. At $m=0$,
the eigenvalue $\gamma_5$ coincides with the helicity so that the
helicity of each quark is also conserved during the scattering.
The process like
\begin{equation}
(R, R)\to (R,L)
\label{mixing}
\end{equation}
with R(L) the right(left) hand helicity will never occur and the
transverse pairing will not couple with the longitudinal one. For
$m\ne 0$, however, the helicity is not the eigenvalue of $\gamma_5$
and is no longer conserved. The two types of pairing do couple via
(\ref{mixing}).

The thermodynamic pressure
\begin{equation}
P=\frac{T}{\Omega}\ln\exp\left(-\frac{H}{T}\right)
\label{pressure}
\end{equation}
with $T$ the temperature and $\Omega$ the volume of the system and the ensemble average of the
operator $O$ is given by
\begin{equation}
<O>=\frac{{\rm
Tr}\left[\exp\left(-\frac{H}{T}\right)O\right]}{{\rm
Tr}\left[\exp\left(-\frac{H}{T}\right)\right]}  \label{average}
\end{equation}
In terms of the plane-wave expansion:
\begin{equation}
\psi=\frac{1}{\sqrt{\Omega}}\sum_{{\bf p},s}(a_{{\bf p},s}u_{{\bf
p},s}e^{i{\bf p}{\bf r}}+b^{+}_{{\bf p},s}v_{{\bf p},s}e^{-i{\bf
p}{\bf r}}) \label{phi}
\end{equation}
with $s(=\pm\frac{1}{2})$ the helicity defined by
\begin{equation}
\sigma\cdot{\bf p} u_{{\bf p},s}=2sp u_{{\bf
p},s}~~~~~~~\sigma\cdot{\bf p} v_{{\bf p},s}=-2sp v_{{\bf
p},s}\label{phii}
\end{equation}
the interaction Hamiltonian reads
\begin{eqnarray}
H_{\rm int}&=&G\bar\psi\gamma_\nu T^l\psi\bar\psi\gamma_\nu T^l\psi\nonumber\\
&=&\frac{1}{\Omega}\sum_{\bf p,p'}a_{{\bf
p'},s_1'}^{\dag}T^la_{{-\bf p},s_2}a_{{-\bf
p'},s_2'}^{\dag}T^la_{{\bf p},s_1}\bar u_{{\bf
p'},s_1'}^{\dag}\gamma_\nu u_{{-\bf p},s_2}\bar u_{{-\bf
p'},s_2'}^{\dag}\gamma_\nu u_{{\bf p},s_1}\label{lag1}\\
&+&\hbox{the terms containing antiquark operators, $b$'s}\nonumber
\end{eqnarray}

The formulae
\begin{equation}
T_{ij}^lT_{km}^l=-\frac{1}{3}(\delta_{ij}\delta_{km}-\delta_{im}\delta_{kj})
+\frac{1}{6}(\delta_{ij}\delta_{km}+\delta_{im}\delta_{kj})
\end{equation}
enables us to decompose the diquark interaction into
color-antisymmetric and symmetric channels and the interaction
within the former is attractive and therefore responsible for
Cooper pairing for $G>0$. We have
\begin{equation}
a_{{\bf p'},s_1'}^{\dag}T^la_{{-\bf p},s_2}a_{{-\bf
p'},s_2'}^{\dag}T^la_{{\bf p},s_1}=\frac{1}{3}a^{\dag}_{{\bf
p'},s_1'}\varepsilon^c\tilde{a}_{{-\bf
p'},s_2'}^{\dag}\tilde{a}_{-{\bf p},s_2}\varepsilon^ca_{{\bf
p},s_1}+\hbox{the color symmetric interaction}
\end{equation}
where the $3\times 3$ antisymmetric matrix $\varepsilon^c$ in
color space is defined by $(\varepsilon^1)=\lambda_5$,
$(\varepsilon^2)=\lambda_7$ and $(\varepsilon^3)=\lambda_2$ and
they coincide with the matrix representation of the angular
momentum operators of spin one with respect to cartesian basis.
We shall designate $J_x$, $J_y$ and $J_z$ for $\lambda_5$, $\lambda_7$
and $\lambda_2$ below. Furthermore, the
gap energy associated to spin-one Cooper pairing is expected much
smaller than the chemical potential. Therefore, we may drop the
antiquark contribution and keep only the color antisymmetric
interaction under the mean field approximation. The relevant
Hamiltonian takes the form
\begin{equation}
H_{\rm eff}=\sum_{\bf p,s}v_F(p-k_F)a^+_{{\bf p}s}a_{{\bf
p}s}-\frac{G}{3\Omega}\sum_{\bf p,p',s_1^\prime,s_2^\prime,s_1,s_2}
A_{s_1^\prime,s_2^\prime;s_1,s_2}(\mathbf{p}',\mathbf{p})a^{\dag}_{{\bf
p'},s_1'}\varepsilon^c\tilde{a}_{{-\bf
p'},s_2'}^{\dag}\tilde{a}_{-{\bf p},s_2}\varepsilon^ca_{{\bf
p},s_1}\label{lag2}
\end{equation}
where
\begin{equation}
A_{s_1^\prime,s_2^\prime;s_1,s_2}(\mathbf{p}',\mathbf{p})\equiv u_{{\bf
p'},s_1'}^{\dag}\gamma_4\gamma_\nu
u_{{-\bf p},s_2}u_{{-\bf p'},s_2'}^{\dag}\gamma_4\gamma_\nu
u_{{\bf p},s_1}
\label{coeff}
\end{equation}
and the approximation $\sqrt{p^2+m^2}-\mu\simeq v_F(p-k_F)$ has
been made with the Fermi momentum $k_F=\sqrt{\mu^2-m^2}$ and the
Fermi velocity $v_F=k_F/\mu$.

To simplify (\ref{lag2}) further, we employ the explicit form of
the four component spinor in the chiral representation
\begin{equation}
u_{{\bf p},s}=\left(
\begin{array}{cc}
\sqrt{\frac{E+2sp}{2E}}\phi_{{\bf p},s} \\
\sqrt{\frac{E-2sp}{2E}}\phi_{{\bf p},s}
\end{array}
\right) \label{helicity}
\end{equation}
where the two component spinor $\phi$ given by
\begin{equation}
\phi_{{\bf p},\frac{1}{2}}=\left(
\begin{array}{cc}
\cos{\frac{\varphi}{2}} \\
e^{i\varphi}\sin{\frac{\theta}{2}}
\end{array}
\right)~~~~~~~~~~~ \phi_{{\bf p},-\frac{1}{2}}=\left(
\begin{array}{cc}
-e^{-i\varphi}\sin{\frac{\theta}{2}}\\
\cos{\frac{\theta}{2}}
\end{array}
\right)\label{helicity2}
\end{equation}
with $(\theta,\varphi)$ the polar angles of $\mathbf{p}$.
Our choice of the phases of $\phi_{{\bf p},\pm\frac{1}{2}}$ is to make them
corresponding to the two columns of the standard Wigner
D-matrix of the angular momentum $J=1/2$, i. e.
\begin{equation}
\left(
\begin{array}{cc}
\phi_{{\bf p},\frac{1}{2}},\phi_{{\bf p},-\frac{1}{2}}
\end{array}
\right)=D^{\frac{1}{2}}(\varphi,\theta,-\varphi)
\end{equation}
where
\begin{equation}
D_{m'm}^J(\alpha,\beta,\gamma)\equiv<Jm'|e^{-iJ_z\alpha}e^{-iJ_y\beta}e^{-iJ_z\gamma}|Jm>
\end{equation}
with {\bf J} the angular momentum operator and
$(\alpha,\beta,\gamma)$ Euler angles. The chiral representation of
gamma matrices is
\begin{equation}
\left(
\begin{array}{cc}
0 & \sigma_{\nu}\\
\bar{\sigma}_{\nu} & 0
\end{array}
\right)
\label{gamma}
\end{equation}
where $\sigma_{\nu}=(1,{\bf\sigma}), \bar\sigma_{\nu}=(1,-{\bf
\sigma})$ with $\sigma$'s the Pauli matrices.

After some algebra detailed in the appendix A, we find that
\begin{equation}
H_{\rm eff}=\sum_{\bf p,s}v_F(p-k_F)a^+_{{\bf p}s}a_{{\bf
p}s}-\frac{4G}{\Omega}\sum_{\bf p,p'}{}^\prime\Phi_\mu^{\nu\dag }({\bf
p'})\Phi_\mu^\nu({\bf p}) \label{Heff}
\end{equation}
where
\begin{equation}
\Phi_\mu^\nu({\bf p})=\sum_{s_1,s_2}
(-1)^{s_2-\frac{1}{2}}e^{-i\theta_{{\bf p}s_2}}B_{s_1s_2}({
p})\left(
\begin{array}{ccc}
\frac{1}{2} & \frac{1}{2}&1\\
-s_2& s_1 & s_2-s_1
\end{array}
\right){D^{1~*}_{\mu,s_2-s_1}}(\varphi,\theta,-\varphi)\tilde{a}_{-{\bf
p}s_2}J^\nu a_{{\bf p}s_1}\label{pree}
\end{equation}
with the phase $\theta_{{\bf p},s}$ is defined by the relation
\begin{equation}
e^{-i\theta_{{\bf p},s}}=-i(-1)^{s-\frac{1}{2}}\phi_{{\bf
p},-s}^{\dag}\phi_{{-\bf p},s}=ie^{2is\varphi}\label{phase}
\end{equation}
and
$B_{\frac{1}{2}\frac{1}{2}}(p)=B_{-\frac{1}{2}-\frac{1}{2}}(p)=\lambda=\frac{m}{E}\simeq\frac{m}{\mu},
B_{\frac{1}{2}-\frac{1}{2}}(p)=B_{-\frac{1}{2}\frac{1}{2}}(p)=1$.
The repeated indexes in the second term of (\ref{HMF}) are summed
over with $\mu,\nu=0,\pm$ and the summation $\sum_{\bf p}^\prime$
extends to half of the momentum space. We have defined
$J_\pm\equiv(\varepsilon^1\pm i\varepsilon^2)$ and $J_0\equiv
\varepsilon^3$ in (\ref{Heff}).

Introducing a long range order $\langle\tilde{a}_{-{\bf
p}s_2}\varepsilon^ca_{{\bf p}s_1}\rangle$ and expanding the
interaction term of (\ref{pree}) to the linear order of the
fluctuation $\tilde{a}_{-{\bf p}s_2}\varepsilon^ca_{{\bf
p}s_1}-\langle\tilde{a}_{-{\bf p}s_2}\varepsilon^ca_{{\bf
p}s_1}\rangle$, we obtain the linearized mean-field Hamiltonian
\begin{eqnarray}
H_{\rm MF.} &=&\sum_{\bf p,s}v_F(p-k_F)a^+_{{\bf p}s}a_{{\bf
p}s}+\frac{9\Omega}{4G}\Delta_\mu^{\nu*}\Delta_\mu^{\nu}-3\sum_{{\bf
p}}[\Delta_\mu^{\nu}\Phi_\mu^\nu({\bf
p})+\Delta_\mu^{\nu*}\Phi_\mu^{\nu\dag}({\bf p})]\label{HMF}
\end{eqnarray}
where the order parameter $\Delta_\mu^\nu$ is defined by
\begin{equation}
\frac{1}{\Omega}\sum_{{\bf p}}\langle\Phi_\mu^\nu({\bf
p})\rangle^*=\frac{3\Delta_\mu^\nu}{4G}
\end{equation}
and will be regarded the element of a $3\times 3$ matrix with
$\mu(\nu)$ the row(column) index. In terms of the Nambu-Gorkov
basis
\begin{equation}
A^+_{\bf p}=\left(
\begin{array}{cccc}
e^{-i\theta_{{\bf p},\frac{1}{2}}}\tilde{a}_{{-\bf p},\frac{1}{2}} &
e^{-i\theta_{{\bf p},-\frac{1}{2}}}\tilde{a}_{{-\bf p},-\frac{1}{2}} & -a^+_{{\bf
p},-\frac{1}{2}}& a^+_{{\bf p},\frac{1}{2}}
\end{array}
\right)~~~~~~A_{\bf p}=\left(
\begin{array}{cccc}
e^{i\theta_{{\bf p},\frac{1}{2}}}\tilde{a}^+_{{-\bf p},\frac{1}{2}}\\
e^{i\theta_{{\bf p},-\frac{1}{2}}}\tilde{a}^+_{{-\bf p},-\frac{1}{2}}\\
 -a_{{\bf p},-\frac{1}{2}}\\
 a_{{\bf p},\frac{1}{2}}
\end{array}
\right)
\end{equation}
the Hamiltonian (\ref{HMF}) takes the form
\begin{equation}
H_{\rm MF.}
=\frac{9}{4G}\Delta_\mu^{\nu*}\Delta_\mu^{\nu}+\sum_{\bf
p}{}^\prime v_F(p-k_F)+\sum_{\bf p}{}^{\prime} A^+_{\bf p}h_{\bf
p}A_{\bf p} \label{lagg}
\end{equation}
where
\begin{equation}
h_{\bf p}=\left(
\begin{array}{cc}
v_F(p-k_F) & M \\
M^{\dag}& -v_F(p-k_F)
\end{array}
\right)\label{hpall}
\end{equation}
and the $6\times 6$ matrix $M$ is defined by
\begin{equation}
M=\sqrt{3}\Delta_\mu^{\nu}J_\nu\times\left(
\begin{array}{cc}
 {D_{\mu,1}^{1*}(\varphi,\theta,-\varphi)} &
 {\frac{\lambda}{\sqrt{2}} D_{\mu,0}^{1*}(\varphi,\theta,-\varphi)} \\
 {\frac{\lambda}{\sqrt{2}} D_{\mu,0}^{1*}(\varphi,\theta,-\varphi)} &
 {D_{\mu,-1}^{1*}(\varphi,\theta,-\varphi)}
\end{array}\right)
\label{sixbysix}
\end{equation}
with $\times$ the direct product and $\lambda=\frac{m}{\mu}$. We have
\begin{equation}
h_{\bf p}^2=\left(
\begin{array}{cc}
v_F^2(p-k_F)^2+MM^{\dag} & 0 \\
0 & -v_F^2(p-k_F)^2-M^{\dag}M
\end{array}
\right)\label{hp2all}
\end{equation}
where both $MM^{\dag}$ and $M^{\dag}M$ have identical nonnegative eigenvalues, $\delta_n^2$ with $n=1,...,6$
and the quasi-particle energy reads $E_{p,n}=\sqrt{v_F^2(p-k_F)^2+\delta_n^2}$.
Replacing the hamiltonian $H$ of (\ref{pressure}) by the linearized one of (\ref{HMF}),
we end up with the pressure under the mean-field approximation
\begin{equation}
 P=-\frac{9}{4G}\Delta_\mu^{\nu*}\Delta_\mu^{\nu}-\frac{1}{\Omega}\sum_{{
p},n}{}^\prime(v_F(p-k_F)-E_{p,n})+\frac{2T}{\Omega}\sum_{{
p},n}{}^\prime\ln\left(1+\exp\left(-\frac{E_{p,n}}{T}\right)\right)
\label{PMF}
\end{equation}
In the massless limit, $\lambda=0$, the off-diagonal elements of the $2\times 2$ matrix in
(\ref{sixbysix}) vanish and we are left with only the transverse pairing.

\section{The thermodynamics of the spin-1 color superconductivity}

The polar, A, planar and CSL are the four canonical phases mostly
discussed in the literature of the spin-1 color superconductivity.
Each of them corresponds to a particular diagonal form of the
$3\times 3$ matrix $\Delta_a^c$ introduced in the last section.
The thermodynamics will be discussed in this section for an
arbitrary quark mass.

To gain more insight to the geometrical structure of these spin-1
phases, we introduce the following two sets of spherical basis
\begin{equation}
\mathbf{e}_\pm\equiv\mp\frac{1}{\sqrt{2}}(\hat{\mathbf{x}}\pm
i\hat{\mathbf{y}}) \qquad\mathbf{e}_0\equiv\hat{\mathbf{z}}
\label{labframe}
\end{equation}
and
\begin{equation}
\mathbf{\epsilon}_\pm\equiv\mp\frac{e^{\pm
i\varphi}}{\sqrt{2}}(\hat{\mathbf{\theta}}\pm
i\hat{\mathbf{\varphi}})\qquad
\mathbf{\epsilon}_0\equiv\hat{\mathbf{p}}\label{bodyframe}
\end{equation}
where $\hat{\mathbf{\theta}}$, $\hat{\mathbf{\varphi}}$ and
$\hat{\mathbf{p}}$ are the unit vectors in the directions of
increasing $\theta$, $\varphi$ and $p$ of the spherical
coordinates of momentum {\bf p}, given by
\begin{eqnarray}
\hat{p}&=&(\sin\theta\cos\varphi,\sin\theta\sin\varphi,\cos\theta)\nonumber\\
\hat{\theta}&=&(\cos\theta\cos\varphi,\cos\theta\sin\varphi,-\sin\theta)\\
\hat{\varphi}&=&(-\sin\varphi,\cos\varphi,0)\nonumber
\end{eqnarray}
The extra phase factor $e^{\pm i\varphi}$ renders $\mathbf{\epsilon}_\pm$ nonsingular at the north pole, $\theta=0$.
It is straightforward to verify that
\begin{equation}
D_{\alpha\beta}^1(\varphi,\theta,-\varphi)=\mathbf{e}_\alpha^*\cdot\mathbf{\epsilon}_\beta
\end{equation}
and the gap matrix takes the compact form
\begin{equation}
M=\sqrt{3}\Delta_\alpha^{\beta}\left(
\begin{array}{cc}
 {(\mathbf{\epsilon}_+)_{\alpha}}^* &
 {\frac{\lambda}{\sqrt{2}}(\mathbf{\epsilon}_0)_{\alpha}}^* \\
 {\frac{\lambda}{\sqrt{2}}(\mathbf{\epsilon}_0)_{\alpha}}^* &
 {(\mathbf{\epsilon}_-)_{\alpha}^*}
\end{array}\right)J_{\beta}
\label{general}
\end{equation}
where the indexes $\alpha$ and $\beta$ run over either the
spherical basis (\ref{labframe}) or cartesian basis
$\hat{\mathbf{x}}$, $\hat{\mathbf{y}}$ and $\hat{\mathbf{z}}$.

Now we are ready to introduce the four canonical spin-1 phases in terms of the circular basis
(\ref{labframe}), with respect to which
\begin{equation}
\vec
J=-\frac{1}{\sqrt{2}}J_-{\mathbf{e}}_++\frac{1}{\sqrt{2}}J_+{\mathbf{e}}_-+J_0{\mathbf{e}}_0
\label{circular}
\end{equation}
with $J_\pm=J_x\pm iJ_y$ and $J_0=J_z$. Each of the canonical
phases corresponds to a particular form of the $3\times 3$ matrix
$\Delta_\alpha^{\beta}$ in (\ref{general})(with $\alpha$ labelling
the rows and $\beta$ the columns). We have
\begin{subequations}
\begin{equation}
\Delta({\rm polar})=\Delta{\rm diag.}(0,0,1)
\end{equation}
\begin{equation}
\Delta({\rm A})=\Delta\left(
\begin{array}{ccc}
0 & 0 & 1\\
0 & 0 & 0\\
0 & 0 & 0
\end{array}\right)
\hbox{   or   }\Delta({\rm A})=\Delta\left(
\begin{array}{ccc}
0 & 0 & 0\\
0 & 0 & 1\\
0 & 0 & 0
\end{array}\right)
\end{equation}
\begin{equation}
\Delta({\rm planar})=\frac{1}{\sqrt{2}}\Delta{\rm diag.}(1,1,0)
\end{equation}
\begin{equation}
\Delta({\rm CSL})=\frac{1}{\sqrt{3}}\Delta{\rm
diag.}(1,1,1)
\end{equation}
\end{subequations}
where $\Delta$ is the gap parameter to be
determined.

Correspondingly, the gap matrix
\begin{equation}
M({\rm polar})=\sqrt{\frac{3}{2}}\Delta\left(
\begin{array}{cc}
 {J_0e^{-i\varphi}\sin\theta} & {J_0\lambda\cos\theta} \\
 {J_0\lambda\cos\theta} & {-J_0e^{i\varphi}\sin\theta}
\end{array}
\right)\label{hpolar}
\end{equation}

\begin{equation}
M({\rm A})=\sqrt{3}\Delta\left(
\begin{array}{cc}
 {-J_0\cos^2\frac{\theta}{2}} & {\frac{\lambda}{2}J_0e^{i\varphi}\sin\theta} \\
 {\frac{\lambda}{2}J_0e^{i\varphi}\sin\theta} &
{-J_0e^{2i\varphi}\sin^2\frac{\theta}{2}}
\end{array}
\right)\label{ha}
\end{equation}

\begin{equation}
M({\rm planar})=\frac{\sqrt{3}}{2}\Delta\left(
\begin{array}{cc}
{-J_-\cos^2\frac{\theta}{2}
+J_+e^{-2i\varphi}\sin^2\frac{\theta}{2}}
& {\frac{\lambda}{2}(J_-e^{i\varphi}+J_+e^{-i\varphi})\sin\theta} \\
{\frac{\lambda}{2}(J_-e^{i\varphi}+J_+e^{-i\varphi})\sin\theta}
& {-J_-e^{2i\varphi}\sin^2\frac{\theta}{2}+J_+\cos^2\frac{\theta}{2}} \\
\end{array}
\right)\label{hplanar1}
\end{equation}
and
\begin{equation}
M_{\rm CSL}=\sqrt{\frac{1}{2}}\Delta\left(
\begin{array}{cc}
-e^{-i\varphi}{\cal J}_- & \lambda{\cal J}_0 \\
\lambda{\cal J}_0 & e^{i\varphi}{\cal J}_+
\end{array}
\right)\label{hcsl1}
\end{equation}
The operators ${\cal J}_\pm$ and ${\cal J}_0$ inside $M_{\rm CSL}$ are defined by
\begin{equation}
{\cal J}_\pm = \mathbf{\epsilon}_{\mp}^*\cdot\mathbf{J} =\pm
e^{\pm i\varphi}(J_\theta\pm iJ_\varphi) \qquad {\cal J}_0 =
\mathbf{\epsilon}_0^*\cdot\mathbf{J}
\end{equation}
with $J_\theta=\hat{\mathbf{\theta}}\cdot\mathbf{J}$ and
$J_\varphi=\hat{\mathbf{\varphi}}\cdot\mathbf{J}$. They satisfy
the same angular momentum algebra as $J_\pm$ and $J_0$ in
(\ref{circular}).

Though the gap matrices (\ref{hpolar}), (\ref{ha}),
(\ref{hplanar1}) and (\ref{hcsl1}) looks complicated, analytical
expressions of the eigenvalues of $MM^{\dag}$ or $M^{\dag}M$ can
be obtained for an arbitrary quark mass. Parametrizing the
eigenvalues by $\Delta^2f^2(\theta)$, we find that
\begin{equation}
f^2(\theta)=\left\{\begin{array}{ll}
\begin{gathered}
(1/8)\left({\sqrt{{\lambda^2}+8}\pm\lambda} \right)^2(d_i=2),\hbox{
}\frac{1}{2}\lambda^2(d_i=2)
\hspace{0.3cm}\hbox{for CSL phase}\\
\end{gathered}
\hfill\\
\begin{gathered}
(3/4)\left(\cos^2\theta+1+\lambda^2{\sin^2\theta}\right)(d_i=4),\hbox{
}0(d_i=2)
\hspace{0.3cm}\hbox{for planar phase}\\
\end{gathered}
\hfill\\
\begin{gathered}
(3/2)\left({\sin^2\theta}+\lambda^2{\cos^2\theta}\right)(d_i=4),\hbox{
}0(d_i=2)
\hspace{0.3cm}\hbox{for polar phase}\\
\end{gathered}
\hfill\\
\begin{gathered}
(3/4)\left(1\pm\sqrt{{\lambda^2}{\sin^2\theta}+{\cos^2\theta}}\right)^2(d_i=2),\hbox{
}0(d_i=2)
\hspace{0.3cm}\hbox{for A phase}\\
\end{gathered}
\end{array}
\right. \label{function}
\end{equation}
where the integer inside the parentheses following each expression indicates
the degeneracy of each distinct eigenvalue. The details of the diagonalization is
shown in Appendix B. The function $f(\theta)$ is $\theta$-dependent for the polar, A and planar
phases and we shall refer to these phases as nonspherical. The CSL phase will be referred to as
spherical because of the constancy of its $f(\theta)$.

Then the pressure corresponding to (\ref{PMF}) becomes:
\begin{eqnarray}
P=-\frac{9}{4G}\Delta^2-\frac{1}{\Omega}\sum_{{
p},i}\frac{d_i}{2}(v_F(p-k_F)-E_{p,i})+\frac{T}{\Omega}\sum_{{
p},i}d_i\ln\left(1+\exp\left(-\frac{E_{p,i}}{T}\right)\right)
\label{pressure0}
\end{eqnarray}
where $E_{{p},i}=\sqrt{v_F^2(p-k_F)^2+\Delta^2f_i^2(\theta)}$. Here
the index $i$ labels the distinct eigenvalues in each line of
(\ref{function}) with $d_i$ the degeneracy. The summation over the
entire momentum space is restored owing to the symmetry of
$f(\theta)$'s under space inversion. Maximizing the pressure with
respect to $\Delta$, we obtain the gap equation
$\left(\frac{\partial P}{\partial\Delta^2}\right)_{T,\mu}=0$, which
determines the temperature dependence of the gap, $\Delta(T)$, up to the transition temperature.

In terms of the parameter
$t=\frac{\Delta(T)}{T}$, the gap equation takes the form
$\ln\frac{\Delta(0)}{\Delta(T)}=\frac{h(t)}{2+\lambda^2}$ with
\begin{eqnarray}
h(t)= \sum_i\frac{d_i}{2}\int_0^{\pi}d\theta\sin\theta
f_i^2(\theta)\int_0^{\infty}dx\frac{1}{\sqrt{x^2+t^2f_i^2(\theta)}[e^{\sqrt{x^2+t^2f_i^2(\theta)}}+1]}
\end{eqnarray}
It follows that
\begin{equation}
T=\frac{\Delta(0)}{t}e^{-\frac{h(t)}{2+\lambda^2}} \label{para1}
\end{equation}
The condensation energy density of the CSC is given by
\begin{equation}
P_s-P_n\equiv\rho_s(t)\frac{\mu^2\Delta_0^2}{2\pi^2}
\label{pdiff}
\end{equation}
with $s$ labeling different pairing states and
$\Delta_0\equiv\Delta_{\rm CSL}(0)$ when $m_s=0$ and we have
\begin{equation}
\rho_s(t)=v_Fe^{-\frac{2}{2+\lambda^2}h(t)}[\frac{2+\lambda^2}{2}+h(t)+2\frac{g(t)}{t^2}-a\frac{\pi^2}{t^2}]
\label{para2}
\end{equation}
with $a=\frac{2}{3}$(1) for nonspherical(spherical) phase and
\begin{equation}
g(t)=\sum_i\frac{d_i}{2}\int_0^{\pi}d\theta\sin\theta\int_0^{\infty}dx\ln[e^{-\sqrt{x^2+t^2f_i^2(\theta)}}+1]
\end{equation}
The curves $P(T)$ may be plotted parametrically according to
(\ref{para1}) and (\ref{para2}) without solving the gap equation
for $T>0$, as we did in \cite{magne}. The transition temperature
$T_c^\lambda$ is determined by (\ref{para1}) in the limit $t\to 0$
with $\Delta(0)$ the solution of the gap equation at $T=0$. We
find that
\begin{equation}
T_c^{\lambda}=(\frac{2K}{\Delta_0})^{(1-\frac{2}{2+\lambda^2}\frac{1}{v_F})}T_c^0\label{temperature}
\end{equation}
where $K$ is a UV cutoff for $\mid{p-k_F}\mid$ in the momentum
integration and is assumed to satisfy the condition
$\Delta_0<<K<<k_F$. We set $K=27$MeV for the numerical calculation
in this paper. For a given mass, $T_c^\lambda$ is universal to all
four phases and the ratio between the pressures of different
phases is independent of the cutoff $K$. This cutoff matters only
when we compare the gaps and pressures of different mass values.
With Fermi velocity $v_F=\sqrt{1-\lambda^2}$, $T_c^{\lambda}$ is a monotonic decreasing function
of $\lambda$ for $0\le\lambda<1$. Then the
transition temperature with massive quarks is always lower than
that in the massless limit. The factor $\rho_s$ vanishes at the
transition temperature $T_c^\lambda$. When $\lambda=0$, the
corresponding curves of $\rho_s(T)$ is the same with what we got
in \cite{magne} in the ultra-relativistic limit. We have $\rho_{\rm
CSL}=1$, $\rho_{\rm planar}=0.98$, $\rho_{\rm polar}=0.88$ and
$\rho_{\rm A}=0.65$ at $T=0$, in agreement with the values reported in
\cite{equality}. In the non-relativistic limit, we get $\rho_{\rm
polar}=\rho_{\rm planar}=2\rho_{\rm A}=\frac{2^{4/3}}{3}\rho_{\rm
CSL}$, consistent with the results in \cite{T}.

\begin{figure}[!htb]
\centering \centering
\includegraphics[scale=0.55, clip=true]{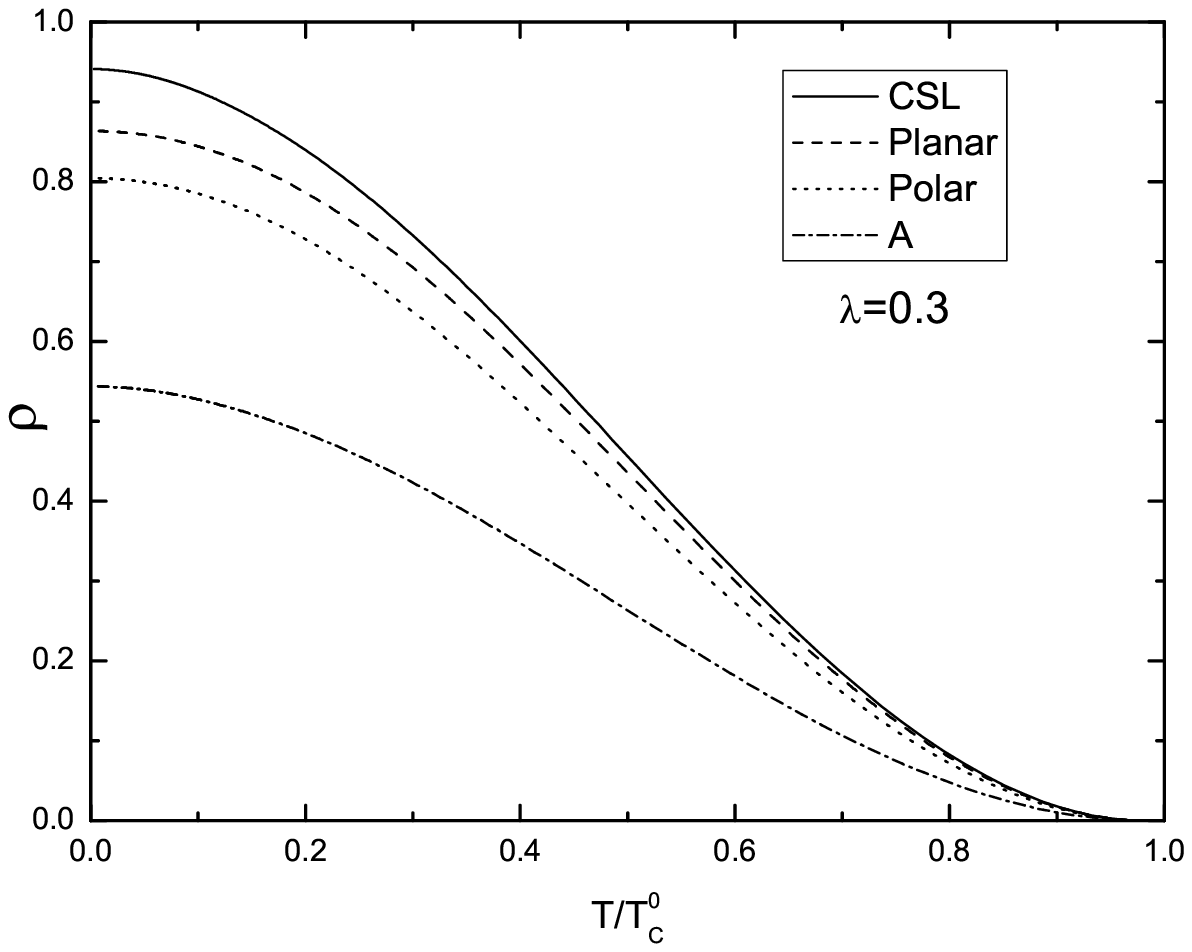}
\hspace{0in}
\includegraphics[scale=0.55, clip=true]{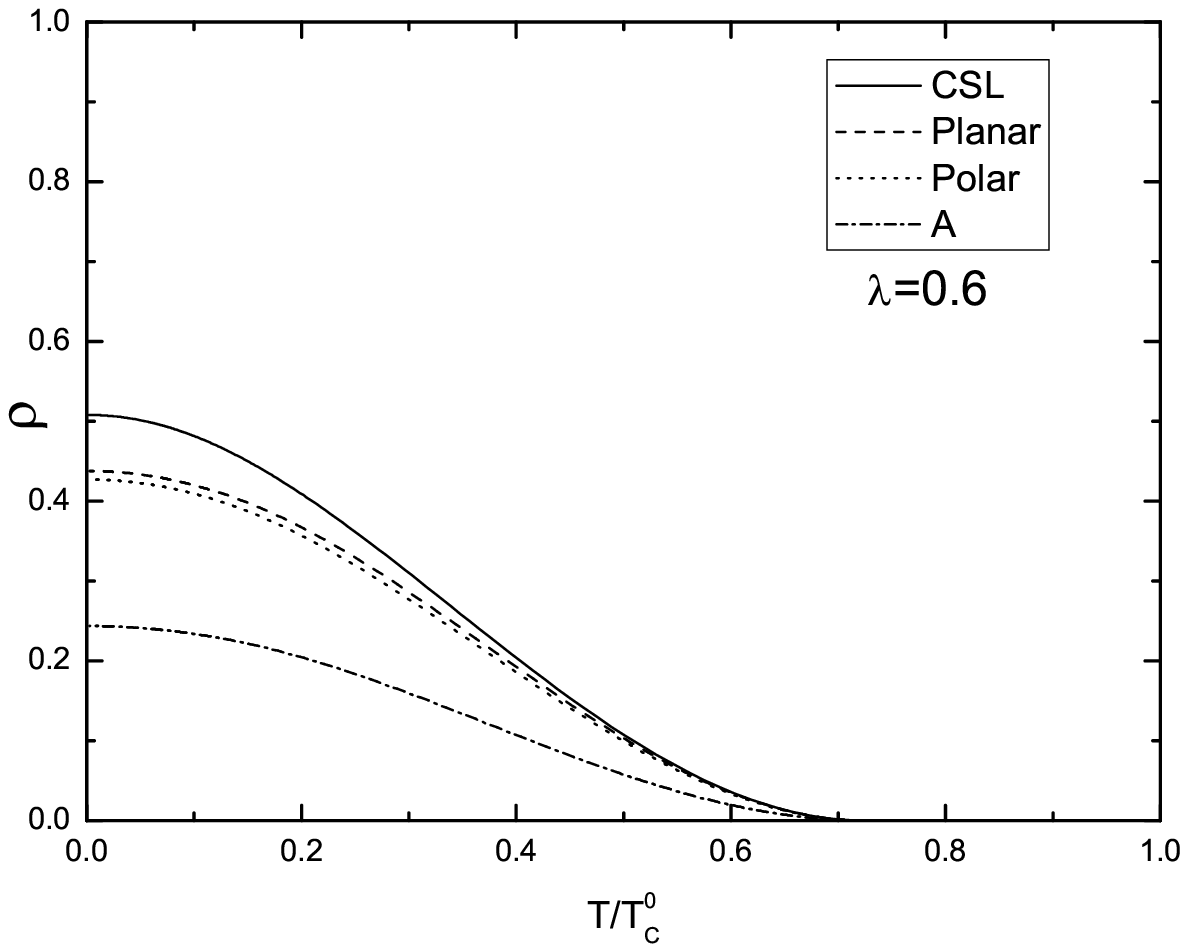}
\caption{The scaled condensation energy dependence on temperature
with different masses $\lambda=0.3$ and
$\lambda=0.6$.\label{fig:phase1}}
\end{figure}

The factor $\rho_s$ versus $T/T_c^\lambda$ is plotted in Fig.1 for
$\lambda=0.3$ and $\lambda=0.6$. For $\mu=500$MeV, the former
corresponds to $m=150$MeV and the latter to $m=300$MeV. The curves
of Fig.\ref{fig:phase1} implies the same inequality
\begin{equation}
P_n<P_{\rm A}<P_{\rm polar}<P_{\rm planar}<P_{\rm CSL}
\label{equa}
\end{equation}
as in the massless limit($\lambda=0$). We expect (\ref{equa}) to hold within the whole domain of $0\le\lambda<1$.

In what follows, we shall identify $\Delta_{\rm CSL}(0)$ with that of the one-gluon
exchange \cite{T,ogluon},
\begin{equation}
\Delta_0=512\pi^4\left(\frac{2}{3}\right)^{\frac{5}{2}}
\frac{\mu}{g^5}\exp\left(-\frac{3\pi^2}{\sqrt{2}g}-\frac{\pi^2+4}{8}-9\right)
\label{onegluon}
\end{equation}
extrapolated to $\mu=500MeV$ and $\alpha_s=\frac{g^2}{4\pi}=1$
with $g$ the QCD running coupling constant. We can obtain the
transition temperature $T_c^{\lambda=0}
=\frac{e^{\gamma_E}}{\pi}\Delta_0$ for u and d quarks in MeV. The
transition temperature of s quarks, of $T_c^{\lambda\neq0}$
follows from (\ref{temperature}). For $K=27$MeV, we find
$T_c^{0.3}=0.98T_c^0$ for $m_s=150$MeV and $T_c^{0.6}=0.683T_c^0$
for $m_s=300$MeV. We should notice that the screening effect
underlying the formulas (\ref{onegluon}) comes from all three
flavors in the massless limit. This inconsistency, however, will
not affect our order of magnitude estimation.

\section{The phase diagram in a magnetic field}

The physics of a quark matter in a magnetic field has received
increasing attention because of the presence of a strong magnetic
field in a compact star or during a noncentral collision of
heavy-ions. The phase structure of 2SC in a magnetic field has
been investigated in \cite{2sc,son}. Equation of state for the CFL
phase in a magnetic field and its implications for compact star
models have been studied in \cite{state}. For an ultra-strong
magnetic field $B$, the spacing of Landau levels becomes
comparable or larger than the quark chemical potential, i.e.
$\sqrt{eB}\geq\mu$, the magnetic field will impact on the pairing
dynamics of CFL \cite{mfirst}. For the typical value of
$\mu$(=500MeV), this requires that $B>10^{18}$G, which may be
implemented inside some magnetar. It was shown in \cite{mfirst}
that an ultra-strong magnetic field may enhance the energy gap of
the CFL for $\sqrt{eB}>>\mu$ and induce a magnetic moment of a
Cooper pair. At a weaker magnetic field, $\sqrt{eB}>>\mu$, a
de-Hass van-Alphen oscillation of the energy gap has been found
\cite{NS,KW}. Alternatively a domain structure may be formed
because of the chiral symmetry breaking and the axial anomaly
\cite{son}. For a spin-1 CSC, in additional to above
possibilities, a magnetic field may offset the balance between the
CSL and nonspherical phases, producing a rich phase structure with
respect to the temperature and the field shown in our previous
work \cite{magne}. This mechanism will be further explored below
taking into account the nonzero mass of strange quarks.

The discussions of proceeding sections imply a nonzero order
parameter
\begin{equation}
\Phi=<\bar\psi_C\Gamma^c\lambda^c\psi> \label{diquark}
\end{equation}
in the coordinate space, where $\psi$ is the quark field,
$\psi_C=\gamma_2\psi^*$ is its charge conjugate, $\lambda^c$ with
$c=2,5,7$ is an antisymmetric Gell-Mann matrices and $\Gamma^c$ is
a $4\times 4$ spinor matrix. We may choose $\Gamma^5=\Gamma^7=0$
for the polar and A phases, $\Gamma^2=0$ for the planar phase but
none of $\Gamma^c$'s vanishes for CSL phase. Depending on the
symmetry of (\ref{diquark}), their responses to an external
magnetic field are quite different.

For CSL phase, the diquark condensate (\ref{diquark}) breaks the
gauge symmetry SU(3)$_c\times$U(1)$_{\rm em}$ completely. But the
Meissner effect for a nonspherical condensate is incomplete,
because it breaks the gauge symmetry partially. Among the
residual gauge group, which leaves the diquark operator inside (\ref{diquark}) unchanged,
there exists a U(1) transformation, $\psi\to
e^{-\frac{i}{2}\lambda_8\theta-iq\phi}\psi$ with $q$ the electric
charge of $\psi$, $\theta=-2\sqrt{3}q\phi$ for the polar and A
phases and $\theta=4\sqrt{3}q\phi$ for the planar phase. The
corresponding gauge field, ${\cal A}_\mu$ is identified with the
electromagnetic field in the condensate and is related to the
electromagnetic field $A$ and the 8-th component of the color
field $A^8$ in the normal phase through a U(1) rotation
\begin{eqnarray}
{\cal A}_\mu &=& A_\mu\cos\gamma-A_\mu^8\sin\gamma\nonumber\\
{\cal V}_\mu &=&
A_\mu\sin\gamma+A_\mu^8\cos\gamma\label{transform}
\end{eqnarray}
where the mixing angle $\gamma$ is given by
$\tan\gamma_{\rm{polar,A}}=2\sqrt{3}q(e/g)$ and
$\tan\gamma_{\rm{planar}}=4\sqrt{3}q(e/g)$ for planar. The 2nd
component of (\ref{transform}) ${\cal V}=0$ because of the
Meissner effect and thereby imposes a constraint inside a
nonspherical CSC, $A_\mu^8=-A_\mu\tan\gamma$, which implies that:
\begin{equation}
{\bf B}^8=-{\bf B}\tan\gamma \label{constraint}
\end{equation}
with ${\bf B}=\mathbf{\nabla}\times\mathbf{A}$.

Expressing the gauge coupling
$\bar\psi\gamma_\mu(eqA_\mu+A_\mu^8\lambda_8/2)\psi$ in terms of
${\cal A}_\mu$ and its orthogonal partner ${\cal V}_\mu$, we
extract the electric charges with respect to ${\cal A}$ in color
space,
\begin{equation}
Q=\left\{\begin{array}{ll}
\begin{gathered}
\frac{3qg}{\sqrt{g^2+12q^2e^2}}{\rm diag.}(0,0,1)
\hspace{0.2cm}\hbox{for polar and A}\\
\end{gathered}
\hfill\\
\begin{gathered}
\frac{3qg}{\sqrt{g^2+48q^2e^2}}{\rm diag.}(1,1,-1)
\hspace{0.2cm}\hbox{for planar}\\
\end{gathered}
\end{array}
\right.\label{charge}
\end{equation}
Because of the nonzero charges of pairing quarks, the planar state
is subject to the impact of Landau orbitals in a magnetic field,
like that for CFL.

The thermal equilibrium in a magnetic field $H\hat{\bf z}$ is
determined by minimizing the Gibbs free energy density,
\begin{equation}
{\cal G}=-P+\frac{1}{2}B^2+\frac{1}{2}\sum_{l=1}^8(B^l)^2-BH
\label{gibbs}
\end{equation}
with respect to $\Delta$, $B$ and $B^l$. Ignoring the induced
magnetization of quarks, the pressure $P$ is given by
(\ref{pressure}), with $\Delta$ given by the solution of the gap
equation. Ignoring the induced
magnetization of quarks, the pressure $P$ is given by
(\ref{pressure}), with $\Delta$ given by the solution of the gap
equation. For a nonspherical CSC pairing, the minimization with
respect to $B$ and $B^l$ is subject to the constraint
(\ref{constraint}). For a hypothetical quark matter of one flavor
only, we find that
\begin{equation}
{\cal G}_{{\rm min.},j}=-P_j-\frac{1}{2}\eta_jH^2 \label{minimum}
\end{equation}
with $j=n$, CSL, polar, A and planar, where $\eta_n=1$, $\eta_{\rm
CSL}=0$ and $\eta_j=\cos^2\gamma_j$ for a nonspherical CSC. The
phase corresponding to minimum among ${\cal G}_{\rm min}$'s above
wins the competition and transition from one phase to another is
first order below $T_c$.

The situation becomes more subtle when quarks of different flavors
coexist even though pairing is within each flavor. Different
electric charges of different quark flavors imply different mixing
angles which may not be compactible with each other. Consider, for instance, a
quark matter of u and d flavors with each flavor in a
non-spherical CSC state with different mixing angles.
(\ref{constraint}) imposes two constraints, which are consistent
with each other only if $B=B^8=0$. Then we end up with an effective
Meissner shielding \cite{SWR}, making it fail to compete with the
phase with both flavors in CSL states. On the other hand, one may
relax the constraints by assuming that the basis underlying the
condensate of u quarks differ from that underlying the condensate
of d quarks by a color rotation. Consequently the constraint
(\ref{constraint}) for each flavor reads $B^8=-B\tan\gamma^u$ and
$B^{\prime8}=-B\tan\gamma^d$. If both flavors stay in the polar or
A phases, which allows ${\bf B}^{1-3}$ to penetrate in, one
may expect that an orthogonal transformation
\begin{eqnarray}
B^{\prime 8}&=&B^8\cos\beta-B^3\sin\beta\nonumber\\
B^{\prime 3}&=&B^8\sin\beta+B^3\cos\beta
\end{eqnarray}
could compromise both constraints. Such a transformation, however,
lies outside the color $SU(3)$ group and therefore, the mutual
rotation of color basis is not an option. The phases of the two
flavor quark matter (u, d) without Meissner effects, which can
compete with (CSL, CSL), include (polar, planar), (polar(normal),
normal(polar)), (A(normal), normal(A)) and (normal, normal).
Notice the coincidence of the mixing angle of the polar state of u
quarks and that of the planar state of d quarks. Also the normal
phase does not impose any constraint on the gauge field and can
coexist with any nonspherical CSC.

The Gibbs free energies remain given by the equations of
(\ref{minimum}), but with $P_n$ and $P_{\rm CSL}$ referring to the
total pressure of all quarks for normal and CSL phases. For
nonspherical phases, $P$ is the total pressure of all flavors with
at least one of them in a nonspherical CSC state and $\gamma$ is
their common mixing angle. For normal-CSC combination, $\gamma$
refers to that of the CSC state. The number of combinations to be
examined is reduced by two criteria: 1) For two combinations of
the same mixing angle, the one with higher pressure wins. 2) For
two combinations of the same pressure, the one with smaller
magnitude of the mixing angle wins. Because the function $\rho_s$
for various CSC phases also satisfy the inequalities (\ref{equa})
up to the transition temperature for an arbitrary mass, it
follows that there are only four phases to be considered in each
case of two and three flavors with nonzero quark masses, which are
shown in Table I.

\begin{table*}
\caption{\label{tab:table1}}
\begin{ruledtabular}
\begin{tabular}{ccccc}
&I&II&III&IV\\
\hline
 2 flavor&$\rm CSL_u, CSL_d$&$\rm (polar)_u, (planar)_d$&$\rm (normal)_u, (polar)_d$&$\rm (normal)_u, (normal)_d$ \\
 3 flavor&$\rm CSL_u, CSL_{d,s}$&$\rm (polar)_u, (planar)_{d,s}$&$\rm (normal)_u, (polar)_{d,s}$&$\rm (normal)_u, (normal)_{d,s}$\\
\end{tabular}
\end{ruledtabular}
\end{table*}

The border between two phases are determined by the equation
\begin{equation}
P_\alpha+\eta_\alpha\frac{H^2}{2}=P_\beta+\eta_\beta\frac{H^2}{2}
\label{border}
\end{equation}
with the subscripts $\alpha$ and $\beta$ labelling the four phases
I-IV.

\begin{figure}[!htb]
\centering
\includegraphics[width=16cm]{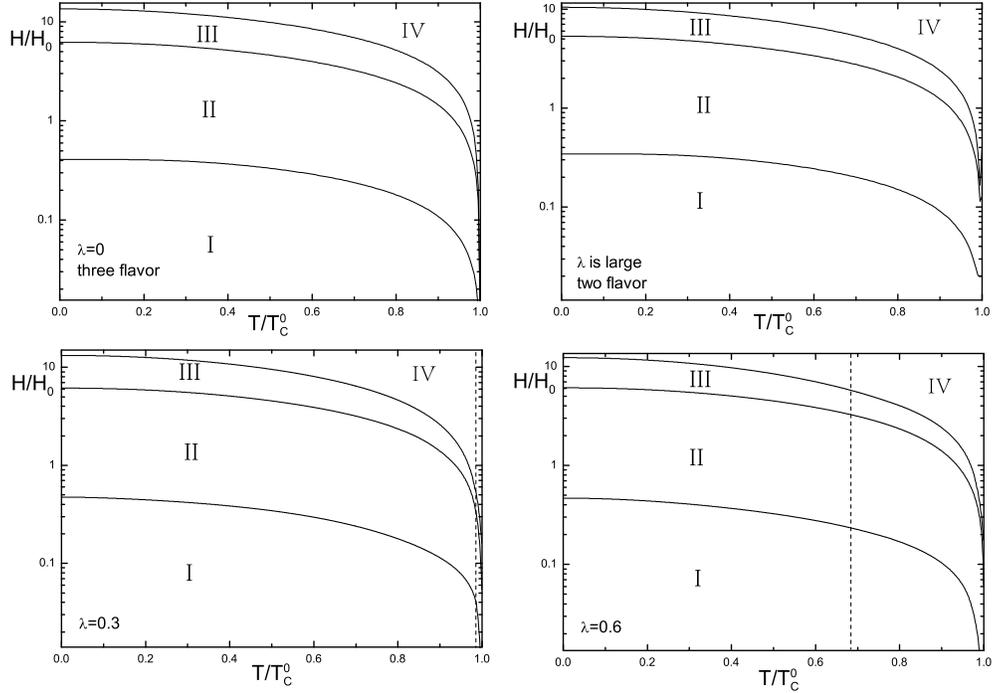}
\caption{H-T phase diagram. These four diagrams corresponding to
the ultra-relativistic limit $m_s=0$, $m_s>>\mu$ and $m_s=150MeV$,
$m_s=300MeV$. \label{Fig:phase1}}
\end{figure}

In a multiflavor quark matter the Fermi momentum of each flavor is
displayed from each other to meet the charge neutrality
requirement (The color neutrality condition is ignored owing to
the small energy gap associated to the single flavor pairing). In
what follows, we shall consider the quark matter of two massless
flavors ($m_u=m_d=0$) and a massive flavor ($m_s\neq0$), coexists
with electrons. Within the mean-field approximation employed in
proceeding sections, the Fermi-momentum displacement can be
determined in the ideal gas limit at zero temperature. The total
pressure under this approximation reads
\begin{equation}
P^{(0)}=-\sum_f E_f-E_e+\mu\sum_f n_f+\mu_q\left(\sum_f q_f
n_f-n_e\right) \label{ideal}
\end{equation}
where $E_f$, $n_f$ and $n_f^q$ are the kinetic energy density and
number density of the quark flavor $f$ with $f=u,d,s$ and
$q_f=(2/3,-1/3,-1/3)$, $E_e$ and $n_e$ are corresponding
quantities for electrons. A charge chemical potential $\mu_q$ is
introduced with the (...) of (\ref{ideal}) the charge number
density. We have
\begin{equation}
E_f=\frac{3}{\pi^2}\int_0^{k_f}dpp^2\sqrt{p^2+m_f^2} \qquad
E_e=\frac{k_e^4}{4\pi^2}
\end{equation}
\begin{equation}
n_f=\frac{1}{\pi^2}k_f^3 \qquad n_e=\frac{1}{3\pi^2}k_e^3
\end{equation}
with $m_f=(0,0,m_s)$. The Fermi momenta, $k_f$ and $k_e$ are
determined by the equilibrium conditions
\begin{equation}
\left(\frac{\partial P^{(0)}}{\partial k_f}\right)_{\mu,\mu_q}
=\left(\frac{\partial P^{(0)}}{\partial k_e}\right)_{\mu,\mu_q}=0
\end{equation}
and the charge neutrality constraint
\begin{equation}
\sum_fq_fn_f-n_e=0
\end{equation}
We find that $k_u=1.001\mu$, $k_d=1.01$ and $k_s=0.941\mu$ for
$m_s=0.3\mu$, and that $k_u=1.004\mu$, $k_d=1.039\mu$ and
$k_s=0.744\mu$ for $m_s=0.6\mu$. We got this H-T diagram,
Fig.{\ref{Fig:phase1}}, and $H_0$ is defined by
\begin{equation}
H_0=\frac{\mu\Delta_0}{\pi}\label{hgb}
\end{equation}

When the strange quark mass is ignored, the transition temperature
of all flavors are the same, this three flavor phase is what we
got in the diagram "$\lambda=0$". The transition from one to
another is first order. When we assume a large mass of $s$ quarks,
$m_s>>\mu$, This two flavor case is certainly unrealistic. When
$m_s$ is comparable to $\mu$, the transition temperature of
strange quark parings is reduced, so when $T_c^{\lambda}<T<T_c^0$,
strange quarks become unpaired, while u and d quark parings
remain. Below $T_c^\lambda$, the condensation energy of strange
quarks rises like $(T_c^\lambda-T)^2$. The transition from three
flavor CSC to two flavor CSC is therefore of second order at zero
magnetic field. Since only the condensation energy, not its
derivatives, enters (\ref{border}), both the phase boundaries and
their slopes with respect to the temperature are continuous at
$T_c^\lambda$. The dashed line in Fig.\ref{Fig:phase1} is the
border between the three flavor region and the two flavor region
in the phase diagram. In the left region of the dashed line, the
corresponding phases combination is "3 flavor" in the
Table.\ref{tab:table1}, and the right region is corresponding to
the upper line "2 flavor". The values of the bounder between two
regions is close to these in massless limit in three flavor, also
these nonspherical phases occupy a significant portion of the H-T
phase diagram for a magnitude of the magnetic field of order
$10^{14}G$. This strong magnetic field is plausible in a compact
star. Due to the not apparent change to the border between the
possible phases, the physical implications of the mass effect to
the cooling behaviors or the latent heat released as the star
cools through the phase boundaries won't be discussed in this
paper, it won't give corrections in order with \cite{heat}. The
smallness of the spin-1 gap makes CSC being of type I and the
critical magnetic field $B\sim\mu\Delta_0<<\mu^2/e$. Therefore,
the magnetic impact on the pairing dynamics as well as the quark
matter magnetization may be neglected \cite{heat}, unlike the
situation considered in \cite{mfirst,NS,KW}.

\section{Concluding remarks}

For the quark matter of moderate density, such as that may reside in
the core of a compact star, the mass of strange quarks, $m_s$, is
not much smaller than  the quark chemical potential  $\mu$ and need
to be taken into account. In this paper, we extend the study of the
four single flavor phases of color superconductivity to include the
effect of the nonzero $m_s$. In spite of the complication of the
coupling between the cross-helicity and equal helicity channels, the
excitation spectrum is obtained analytically. We have explored its
correction to the pressure and the transition temperature
numerically. It is found that mass effect reduces the pressure and
transition temperature of strange quarks, but it doesn't change the
ranking $P_n<P_{\rm A}<P_{\rm polar}<P_{\rm planar}<P_{\rm CSL}$ of
the pressure for the four canonical single flavor phases for the
values of $m_s$ we examined. We suspect that the above inequality
holds for an arbitrary $m_s$. Then we generalized the previous work
to the quark matter with massless u and d flavors and the massive s
flavor. Because the transition temperature for strange quark parings
is lower than that of massless quark pairings, this new H-T diagram
consists of the two flavor CSC for $T_c^{\lambda}< T < T_c^0$ and
the three flavor CSC for $0 < T <T_c^{\lambda}$. The three flavor
and the two flavor region will be occupied by the same phases I-IV
in Table I with the same relative positions.  There is a second
order phase transition from three flavor condensate to two flavor
condensate at transition temperature $T_c^{\lambda}$ where strange
quarks condense. The phase boundaries in the two regions join
smoothly. As an order of magnitude estimate, we calibrated our model
$T_c$ against the that from the QCD one-gluon exchange in the chiral
limit and found the typical magnitude of the magnetic field in the
phase diagram falls within the range of the plausible magnetic field
inside a compact star in the literature.

On the other hand, the effective Lagrangian (\ref{lag}) we
employed in this paper is by no means the most general one. The
Lorentz covariance of (\ref{lag}) is unlikely in a medium and the
coupling $G$ may depends on $m_s$. Taking the one-gluon exchange
as a reference, the screenings of the color magnetic channel and
the color electric one by the medium are very different and should
depends the masses of quarks. These properties is likely to
persists qualitatively at the moderate density and should be
reflected in the effective action to some extents. Also, the
inequality (\ref{equa}) may not be as robust as people thought. A
purely Ginzburg-Landau analysis \cite{B} reveals some parameter
region where the ranking (\ref{equa}) is offset even without a
magnetic field. The microscopic mechanism supporting this
observation, based on the most general four-fermion effective
action or others, remains to be unveiled.

Throughout this paper, we take the massless limit of $u$ and $d$
flavors and then gapless excitations exist in all I-IV phases of
Table.I. In reality, the chiral restoration transition from low
density to high density may be a crossover and $u$ and $d$ quarks
may also acquire nonzero masses from the chiral condensate.
Consequently, the excitations of the phase I where all flavors are
in CSL will be gapped. This is welcome since it will slow down the
direct Urca processes of cooling in a compact star by spin-1 CSC
alone \cite{blaschke}. But the gapless modes remains for
nonspherical states and phase diagrams Fig.2 are still valid
qualitatively. Therefore the magnetic field inside the star cannot
exceed the magnitude along the border line between I and II of
Fig.2 for a slow cooling process.

\begin{acknowledgments}

The authors are grateful to Thomas Sch$\ddot{a}$fer for a
communication on the NR limit. We would like to extend our
gratitude to  Bo Feng for useful discussions. The work
of D. F. H. and H. C. R. is supported in part by NSFC under grant
Nos. 10575043, 10735040,11135011.
\end{acknowledgments}

\appendix
\section{}

In this appendix, we shall fill in the details from (\ref{lag2}) to (\ref{Heff}).
Substituting (\ref{helicity}) and (\ref{gamma}) into (\ref{coeff}), we obtain
that
\begin{eqnarray}
A_{s_1^\prime,s_2^\prime;s_1,s_2}(\mathbf{p}',\mathbf{p})
&=&2C_{s'_1s'_2}({\bf p}')C_{-s_2-s_1}({\bf p})\phi_{{\bf
p'},s_1'}^{\dag}\bar{\sigma}^{\nu}\phi_{{-\bf p},s_2}\phi_{{-\bf
p'},s_2'}^{\dag}\bar{\sigma}^{\nu}\phi_{{\bf p},s_1}\nonumber\\
&+&2C_{-s'_1-s'_2}({\bf p}')C_{s_2s_1}({\bf p})\phi_{{\bf
p'},s_1'}^{\dag}{\sigma}^{\nu}\phi_{{-\bf p},s_2}\phi_{{-\bf
p'},s_2'}^{\dag}{\sigma}^{\nu}\phi_{{\bf p},s_1}\\
&+&2C_{s'_1-s'_2}({\bf p}')C_{-s_2s_1}({\bf p})\phi_{{\bf
p'},s_1'}^{\dag}\bar{\sigma}^{\nu}\phi_{{-\bf p},s_2}\phi_{{-\bf
p'},s_2'}^{\dag}{\sigma}^{\nu}\phi_{{\bf p},s_1}\nonumber\\
&+&2C_{-s'_1s'_2}({\bf p}')C_{s_2-s_1}({\bf p})\phi_{{\bf
p'},s_1'}^{\dag}{\sigma}^{\nu}\phi_{{-\bf p},s_2}\phi_{{-\bf
p'},s_2'}^{\dag}\bar{\sigma}^{\nu}\phi_{{\bf p},s_1}\nonumber
\end{eqnarray}
where $C_{ss'}(p)=\frac{\sqrt{(E+2sp)(E+2s'p)}}{2E}$
It follows from the identity
\begin{equation}
(\sigma_j)_{\alpha\beta}(\sigma_j)_{\gamma\delta}
=2\delta_{\alpha\delta}\delta_{\beta\gamma}
-\delta_{\alpha\beta}\delta_{\gamma\delta}
\end{equation}
that
\begin{eqnarray}
\phi_{{\bf p'},s_1'}^{\dag}\bar{\sigma}^{\nu}\phi_{{-\bf
p},s_2}\phi_{{-\bf p'},s_2'}^{\dag}\bar{\sigma}^{\nu}\phi_{{\bf
p},s_1}&=&\phi_{{\bf p'},s_1'}^{\dag}{\sigma}^{\nu}\phi_{{\bf
p},s_2}\phi_{{-\bf p'},s_2'}^{\dag}{\sigma}^{\nu}\phi_{{-\bf
p},s_1}\nonumber\\&=&2\phi_{{\bf p'},s_1'}^{\dag}\phi_{{-\bf
p},s_2}\phi_{{-\bf p'},s_2'}^{\dag}\phi_{{\bf p},s_1}-2\phi_{{\bf
p'},s_1'}^{\dag}\phi_{{\bf p},s_1}\phi_{{-\bf
p'},s_2'}^{\dag}\phi_{{-\bf p},s_2}\\
\phi_{{\bf p'},s_1'}^{\dag}\bar{\sigma}^{\nu}\phi_{{-\bf
p},s_2}\phi_{{-\bf p'},s_2'}^{\dag}{\sigma}^{\nu}\phi_{{\bf
p},s_1}&=&\phi_{{\bf p'},s_1'}^{\dag}{\sigma}^{\nu}\phi_{{-\bf
p},s_2}\phi_{{-\bf
p'},s_2'}^{\dag}\bar{\sigma}^{\nu}\phi_{{\bf p},s_1}\nonumber\\
&=&2\phi_{{\bf p'},s_1'}^{\dag}\phi_{{\bf p},s_1}\phi_{{-\bf
p'},s_2'}^{\dag}\phi_{{-\bf p},s_2}\nonumber
\end{eqnarray}
For two different momenta, ${\bf p}$ and ${\bf p'}$, we have
\begin{equation}
\phi_{{\bf p'},s'}^{\dag}\phi_{{\bf
p},s}=(D^{\frac{1}{2}\dag}(\varphi',\theta',-\varphi')D^{\frac{1}{2}}(\varphi,\theta,-\varphi))_{s's}
=D^{\frac{1}{2}}_{s's}(R)\label{helicity5}
\end{equation}
where $R$ stands for the Euler angles corresponding to the product
of the rotations specified by $(\varphi, \theta, -\varphi)$ and
$(\varphi', \theta', -\varphi')$. Together with the orthonormal
relation $\phi_{{\bf p},s}^{\dag}\phi_{{\bf p},s'}=\delta_{ss'}$,
we obtain that
\begin{eqnarray}
&{}&A_{s_1^\prime,s_2^\prime;s_1,s_2}(\mathbf{p}',\mathbf{p})\nonumber\\
&=&4(-1)^{s_2+s_2'-1}e^{i(-\theta_{{\bf p}s_2}+\theta_{{\bf
p}'s'_2})}C_{s'_1s'_2}( p')C_{-s_2-s_1}(p)e^{i(-\theta_{{\bf
p}-s_2}+\theta_{{\bf
p}'-s'_2})}(D^{\frac{1}{2}}_{s'_1-s_2}(R)D^{\frac{1}{2}}_{-s'_2s_1}(R)-D^{\frac{1}{2}}_{s'_1s_1}(R)D^{\frac{1}{2}}_{-s'_2-s_2}(R))\label{1111}\nonumber\\
&+&4(-1)^{s_2+s_2'-1}C_{s'_1-s'_2}(p')C_{s_1-s_2}(p)e^{i(-\theta_{{\bf
p}s_2}+\theta_{{\bf
p}'s'_2})}D^{\frac{1}{2}}_{s'_1-s_2}(R)D^{\frac{1}{2}}_{-s'_2s_1}(R)\label{2}
\end{eqnarray}
where the phase $\theta_{\mathbf{p,s}}$ is defined in
(\ref{phase}) and satisfies the relation
\begin{equation}
e^{i\theta_{-\mathbf{p},s}}=-e^{i\theta_{\mathbf{p},s}}
\label{inversion}
\end{equation}
Because
\begin{equation}
D^{\frac{1}{2}}_{s'_1-s_2}(R)D^{\frac{1}{2}}_{-s'_2s_1}(R)-D^{\frac{1}{2}}_{s'_1s_1}(R)D^{\frac{1}{2}}_{-s'_2-s_2}(R))
=detD^{\frac{1}{2}}(R)\epsilon_{s'_1-s'_2}\epsilon_{s_1-s_2}=\epsilon_{s'_1-s'_2}\epsilon_{s_1-s_2}
\end{equation}
in (\ref{1111}), $\epsilon_{s'_1-s'_2}\epsilon_{s_1-s_2}\neq0$
requires that $s'_1=s'_2,s_1=s_2$. Then the diquark operator of
equal helicity is even in ${\bf p}$ because of the equation
(\ref{phase}), so sum over ${\bf p}$ will make it vanish on
account of (\ref{inversion}). So this part doesn't contribute.

Using the formula of Wigner D-functions
\begin{equation}
D^A_{aa'}(\alpha,\beta,\gamma)D^B_{bb'}(\alpha,\beta,\gamma)=\sum_C(2C+1)\left(
\begin{array}{ccc}
A & B &C \\
a & b &c
\end{array}
\right)\left(
\begin{array}{ccc}
A & B &C \\
a' & b' &c'
\end{array}
\right)D^{C*}_{cc'}(\alpha,\beta,\gamma)\label{helicity2}
\end{equation}
we find:
\begin{eqnarray}
D^{\frac{1}{2}}_{s'_1-s_2}(R)D^{\frac{1}{2}}_{-s'_2s_1}(R)&=&3
\left(
\begin{array}{ccc}
\frac{1}{2} & \frac{1}{2}&1\\
s'_1& -s'_2 & s'_2-s'_1
\end{array}
\right)\left(
\begin{array}{ccc}
\frac{1}{2} & \frac{1}{2}&1\\
-s_2& s_1 & s_2-s_1
\end{array}
\right) {D^{1*}_{s'_2-s'_1,s_2-s_1}}(R)\nonumber\\
&+&\left(
\begin{array}{ccc}
\frac{1}{2} & \frac{1}{2}&0\\
s'_1& -s'_2 & s'_2-s'_1
\end{array}
\right)\left(
\begin{array}{ccc}
\frac{1}{2} & \frac{1}{2}&0\\
-s_2& s_1 & s_2-s_1
\end{array}
\right) {D^{0*}_{s'_2-s'_1,s_2-s_1}}(R)
\end{eqnarray}
where $s'_1, s'_2, s_1, s_2$ can take values of $\pm\frac{1}{2}$.
The ${D^{0*}_{s'_2-s'_1,s_2-s_1}}(R)$ part  doesn't contribute
either, because it also requires $s'_1=s'_2,s_1=s_2$, which makes
it vanish when to sum over ${\bf p}$.

It follows from the second equality of (\ref{helicity5}) that
\begin{equation}
D^1_{m_1m_2}(R)=\sum_m
D^{1*}_{mm_1}(\varphi',\theta',-\varphi')D^1_{mm_2}(\varphi,\theta,-\varphi)
\end{equation}
and we arrive at
\begin{eqnarray}
A_{s_1^\prime,s_2^\prime;s_1,s_2}(\mathbf{p}',\mathbf{p})
&=&12(-1)^{s_2+s_2'-1}C_{s'_1-s'_2}(p')C_{s_1-s_2}(p)e^{i(-\theta_{{\bf
p}s_2}+\theta_{{\bf p}'s'_2})} \left(
\begin{array}{ccc}
\frac{1}{2} & \frac{1}{2}&1\\
s'_1& -s'_2 & s'_2-s'_1
\end{array}
\right)\left(
\begin{array}{ccc}
\frac{1}{2} & \frac{1}{2}&1\\
-s_2& s_1 & s_2-s_1
\end{array}
\right)\nonumber\\
&\times&\sum_m{D^1_{m,s'_2-s'_1}}(R){D^1_{m,s_2-s_1}}(R)+...
\label{coeff1}
\end{eqnarray}
with "..." the terms not contributing to the summation over
momenta.

Substituting (\ref{coeff1}) into (\ref{lag2}), the interaction term of (\ref{lag2}) becomes
\begin{equation}
\sum_{\bf p,\bf p',s_1^\prime,s_2^\prime,s_1,s_2}
A_{s_1^\prime,s_2^\prime;s_1,s_2}(\mathbf{p}',\mathbf{p})a^{\dag}_{{\bf
p'},s_1'}\varepsilon^c\tilde{a}_{{-\bf
p'},s_2'}^{\dag}\tilde{a}_{-{\bf p},s_2}\varepsilon^ca_{{\bf
p},s_1}
=12\sum_{\bf p,\bf p'}\Xi_\mu^{\nu\dag }({\bf
p'})\Xi_\mu^\nu({\bf p})
\end{equation}
with
\begin{equation}
\Xi_\mu^\nu({\bf p})=\sum_{s_1,s_2}
(-1)^{s_2-\frac{1}{2}}e^{-i\theta_{{\bf p}s_2}}C_{s_1-s_2}({
p})\left(
\begin{array}{ccc}
\frac{1}{2} & \frac{1}{2}&1\\
-s_2& s_1 & s_2-s_1
\end{array}
\right){D^{1~*}_{\mu,s_2-s_1}}(\varphi,\theta,-\varphi)\tilde{a}_{-{\bf
p}s_2}J^\nu a_{{\bf p}s_1}\label{Xi}
\end{equation}
If follows from the explicit form of the phase factor (\ref{phase})
and the symmetry properties of the D-functions that
\begin{equation}
\Xi_\mu^\nu(-{\bf p})=\sum_{s_1,s_2}
(-1)^{s_2-\frac{1}{2}}e^{-i\theta_{{\bf p}s_2}}C_{-s_1s_2}({
p})\left(
\begin{array}{ccc}
\frac{1}{2} & \frac{1}{2}&1\\
-s_2& s_1 & s_2-s_1
\end{array}
\right){D^{1~*}_{\mu,s_2-s_1}}(\varphi,\theta,-\varphi)\tilde{a}_{-{\bf
p}s_2}J^\nu a_{{\bf p}s_1}\label{Xi1}
\end{equation}
On writing $\sum_{\bf p,\bf p'}^\prime\Xi_\mu^{\nu\dag}({\bf p})$ with the summation $\sum^\prime$ extending half space of
and we end up with (\ref{Heff}) with
\begin{equation}
\Phi_\mu^\nu(\mathbf{p})\equiv\Xi_\mu^\nu(\mathbf{p})+\Xi_\mu^\nu(-\mathbf{p})
\end{equation}
given by (\ref{pree}).

\section{}

In this section, we will give the details of the diagonalization
procedure of the $6\times 6$ matrix $MM^\dag$ for each single flavor phase.
We shall write $MM^\dag\equiv \Delta^2 {\cal M}$. The eigenvalues of ${\cal M}$
corresponds to $f^2(\theta)$ shown in (\ref{function}).

{\noindent \it{The polar phase:}}

It's straightforward to show
 \begin{equation}
{\cal M}_{\rm polar}=\frac{3}{2}\left(
\begin{array}{cc}
(\sin^2\theta+\lambda^2\cos^2\theta)J_0^2 & 0 \\
0& (sin^2\theta+\lambda^2\cos^2\theta)J_0^2
\end{array}
\right)\label{mpolar}
\end{equation}
and the color operator $J_0^2$ decouples.
The eigenvalues of $J_0^2$ are $1, 1, 0$ and the functional forms of $f(\theta)$ are therefore given by the 3rd
line of (\ref{function}).

{\noindent \it{The A phase:}}

In this case we have
 \begin{equation}
{\cal M}_{\rm A}=3\left(
\begin{array}{cc}
(2\sin^4\frac{\theta}{2}+\frac{1}{2}\lambda^2\sin^2\theta)J_0^2 & -\lambda\sin\theta e^{i\varphi}J_0^2 \\
-\lambda\sin\theta e^{i\varphi}J_0^2 &
(2\cos^4\frac{\theta}{2}+\frac{1}{2}\lambda^2\sin^2\theta)J_0^2
\end{array}
\right)\label{mA}
\end{equation}

The color operator $J_0^2$, which has eigenvalues 1,1 and 0,
decouples again. The forms of $f(\theta)$ given by the fourth line
of (\ref{function}) correspond to the eigenvalues of the $2\times
2$ matrix obtained from (\ref{mA}) by replacing $J_0^2$ with its
eigenvalues.

{\noindent \it{The planar phase:}}

The diagonalization of $MM^{\dag}$ is less straightforward because of the coupling between the helicity
and the color indexes. In terms of $J_\pm^\prime\equiv J_\pm e^{\mp i\varphi}$, we have
\begin{equation}
{\cal M}_{\rm Planar}=\frac{3}{4}\left(
\begin{array}{cc}
a & b\\
c & d
\end{array}
\right)
\end{equation}
where:
\begin{subequations}
{\small
\begin{eqnarray}
a&=&\left(\cos^4{\frac{\theta}{2}}+\frac{1}{4}\lambda^2\sin^2\theta\right)J_-^\prime J_+^\prime
+\left(\sin^4{\frac{\theta}{2}}+\frac{1}{4}\lambda^2\sin^2\theta\right)J_+^\prime J_-^\prime
-\frac{1}{4}(1-\lambda^2)(J_-^{\prime2}+J_+^{\prime2})\sin^2\theta\\
b&=&\frac{\lambda}{2}\sin{\theta}e^{-i\varphi}[J_+^\prime,J_-^\prime]~~~~~~ c=b^\dag\\
d&=&\left(\sin^4{\frac{\theta}{2}}+\frac{1}{4}\lambda^2\sin^2\frac{\theta}{2}\right)J_-^\prime
J_+^\prime
+\left(\cos^4{\frac{\theta}{2}}+\frac{1}{4}\lambda^2\sin^2\theta\right)J_+^\prime J_-^\prime
-\frac{1}{4}(1-\lambda^2)(J_-^{\prime2}+J_+^{\prime2})\sin^2\theta
\end{eqnarray}}
\end{subequations}
Since $J_\pm^\prime\equiv J_\pm e^{\mp
i\varphi}$ and $J_0$ satisfy the same angular momentum algebra as
$J_\pm$ and $J_0$ do, we shall work in the representation where
\begin{equation}
J_+^\prime=\sqrt{2}\left(
\begin{array}{ccc}
0 & 1 & 0\\
0 & 0 & 1\\
0 & 0 & 0
\end{array}
\right)~~~~~~ J_-^\prime=\sqrt{2}\left(
\begin{array}{ccc}
0 & 0 & 0 \\
1 & 0 & 0 \\
0 & 1 & 0
\end{array}
\right).
\end{equation}
and $J_0={\rm diag}(1,0,-1)$. It follows that
\begin{subequations}
\begin{equation}
a=\left(
\begin{array}{ccc}
2\sin^4\frac{\theta}{2}+\frac{\lambda^2}{2}\sin^2\theta & 0 &\frac{1}{2}(1-\lambda^2)\sin^2\theta\\
0 & 1+\cos^2{\theta}+\lambda^2\sin^2{\theta} & 0\\
\frac{1}{2}(1-\lambda^2)\sin^2\theta & 0 &
2\cos^4\frac{\theta}{2}+\frac{\lambda^2}{2}\sin^2\theta
\end{array}
\right)
\end{equation}
\begin{equation}
b=\lambda\sin\theta e^{-i\varphi}\left(
\begin{array}{ccc}
1 & 0 & 0\\
0 & 0 & 0\\
0 & 0 &-1
\end{array}
\right)
\end{equation}
\begin{equation}
d=\left(
\begin{array}{ccc}
2\cos^4\frac{\theta}{2}+\frac{\lambda^2}{2}\sin^2\theta & 0 &\frac{1}{2}(1-\lambda^2)\sin^2\theta\\
0 & 1+\cos^2{\theta}+\lambda^2\sin^2{\theta} & 0\\
\frac{1}{2}(1-\lambda^2)\sin^2\theta & 0 &
2\sin^4\frac{\theta}{2}+\frac{\lambda^2}{2}\sin^2\theta
\end{array}
\right)
\end{equation}
\end{subequations}
By permutations of the rows and columns, this 6
by 6 matrix is transformed into the block-diagonal form
\begin{equation}
{\cal M}_{\rm Planar}=\left(
\begin{array}{ccc}
1+\cos^2{\theta}+\lambda^2\sin^2{\theta}&0&0\\
0&1+\cos^2{\theta}+\lambda^2\sin^2{\theta}&0\\
0&0&M_4
\end{array}
\right)
\label{planarM}
\end{equation}
where $M_4$ is a 4 by 4 matrix, given by
\begin{equation}
M_4=\left(
\begin{array}{cccc}
2\sin^4\frac{\theta}{2}+\frac{\lambda^2}{2}\sin^2\theta & \frac{1}{2}(1-\lambda^2)\sin^2\theta & \lambda\sin\theta e^{-i\varphi} & 0\\
\frac{1}{2}(1-\lambda^2)\sin^2\theta & 2\cos^4\frac{\theta}{2}+\frac{\lambda^2}{2}\sin^2\theta & 0 & -\lambda\sin\theta e^{-i\varphi}\\
\lambda\sin\theta e^{i\varphi} & 0 & 2\cos^4\frac{\theta}{2}+\frac{\lambda^2}{2}\sin^2\theta & \frac{1}{2}(1-\lambda^2)\sin^2\theta\\
0 & \lambda\sin\theta e^{i\varphi} &
\frac{1}{2}(1-\lambda^2)\sin^2\theta &
2\sin^4\frac{\theta}{2}+\frac{\lambda^2}{2}\sin^2\theta
\end{array}
\right)
\end{equation}
It is straightforward to show that the secular equation
\begin{equation}
{\rm det}(M_4-z)=z^2(z-1-\cos^2\theta-\lambda^2\sin^2\theta)^2
\end{equation}
which, together with (\ref{planarM}), yield the eigenvalues in the
second line of (\ref{function}).

{\noindent \it{The CSL phase:}}

In terms of the operator ${\cal J}_\pm$ and ${\cal J}_0$, the matrix
${\cal M}$ of CSL takes the form
\begin{equation}
{\cal M}_{\rm CSL}=\frac{1}{2}\left(
\begin{array}{cc}
{\cal J}_-{\cal J}_++\lambda^2{\cal J}_0^2 & \lambda e^{-i\varphi}[{\cal J}_0, {\cal J}_-] \\
-\lambda e^{i\varphi}[{\cal J}_0,{\cal J}_+]  & {\cal J}_+{\cal
J}_-+\lambda^2{\cal J}_0^2
\end{array}\right)
\label{cslm}
\end{equation}
The operators ${\cal J}_\pm$ and ${\cal J}_0$ satisfy the same
algebraic relations as $J_\pm$ and $J_0$, such as $[{\cal J}_0,
{\cal J}_\pm]=\pm{\cal J}_\pm$. In the representation where ${\cal
J}_0$ is diagonal, i.e. ${\cal J}_0={\rm diag.}(1,0,-1)$,
\begin{equation}
{\cal J}_+ = \sqrt{2}\left(
\begin{array}{ccc}
0&1&0\\
0&0&1\\
0&0&0
\end{array}\right)\hbox{     and     }
{\cal J}_- = \sqrt{2}\left(
\begin{array}{ccc}
0&0&0\\
1&0&0\\
0&1&0
\end{array}\right)
\end{equation}
 we have
${\cal J}_-{\cal J}_+={\rm diag}(0,2,2)$, ${\cal J}_+{\cal J}_-={\rm diag}(2,2,0)$.
Substituting these into (\ref{cslm}), we find that
\begin{equation}
{\cal M}_{\rm CSL} = \frac{1}{2}\left(
\begin{array}{cccccc}
\lambda^2&0&0&0&0&0\\
0&2&0&-\sqrt{2}\lambda e^{-i\varphi}&0&0\\
0&0&2+\lambda^2&0&-\sqrt{2}e^{-i\varphi}&0\\
0&-\sqrt{2}\lambda e^{i\varphi}&0&2+\lambda^2&0&0\\
0&0&-\sqrt{2}\lambda e^{i\varphi}&0&2&0\\
0&0&0&0&0&\lambda^2
\end{array}
\right)\label{cslnumber}
\end{equation}
By permutations of the rows and columns, this 6 by 6 matrix is transformed into the
block-diagonal form
\begin{equation}
\frac{1}{2}\left(
\begin{array}{cccccc}
\lambda^2&0&0&0&0&0\\
0&\lambda^2&0&0&0&0\\
0&0&2&-\sqrt{2}\lambda e^{-i\varphi}&0&0\\
0&0&-\sqrt{2}\lambda e^{i\varphi}&2+\lambda^2&0&0\\
0&0&0&0&2+\lambda^2&-\sqrt{2}\lambda e^{-i\varphi}\\
0&0&0&0&-\sqrt{2}\lambda e^{i\varphi}&2\\
\end{array}
\right)\label{cslnumber}
\end{equation}
and the eigenvalues in the first line of (\ref{function}) follow then.

\end{document}